# Gallium transformation under femtosecond laser excitation:
# Phase coexistence and incomplete melting

O. P. Uteza[1,2], E. G. Gamaly[1]*, A. V. Rode[1], M. Samoc[1], B. Luther-Davies[1]

*[1]Laser Physics Centre, Research School of Physical Sciences and Engineering*
*Australian National University, Canberra, ACT 0200 Australia*
*[2]LP3 - Lasers, Plasmas and Photonic Processes Laboratory,*
*FRE 2165 CNRS – Aix-Marseille II University,*
*Case 917, 163, Avenue de Luminy, 13288 Marseille Cedex 9, France*

## Abstract

The reversible phase transition induced by femtosecond laser excitation of Gallium has been studied by measuring the dielectric function at 775 nm with ~ 200 fs temporal resolution. The real and imaginary parts of the transient dielectric function were calculated from absolute reflectivity of Gallium layer measured at two different angles of incidence, using Fresnel formulas. The time-dependent electron-phonon effective collision frequency, the heat conduction coefficient and the volume fraction of a new phase were restored directly from the experimental data, and the time and space dependent electron and lattice temperatures in the layer undergoing phase transition were reconstructed without *ad hoc* assumptions. We converted the temporal dependence of the electron-phonon collision rate into the temperature dependence, and demonstrated, for the first time, that the electron-phonon collision rate has a non-linear character. This temperature dependence converges into the known equilibrium function during the cooling stage. The maximum fraction of a new phase in the laser-excited Gallium layer reached only 60% even when the deposited energy was two times the equilibrium enthalpy of melting. We have also demonstrated that the phase transition pace and a fraction of the transformed material depended strongly on the thickness of the laser-excited Gallium layer, which was of the order of several tens of nanometers for the whole range of the pump laser fluencies up to the damage threshold. The kinetics of the phase transformation after the laser excitation can be understood on the basis of the classical theory of the first-order phase transition while the duration of non-thermal stage appears to be comparable to the sub-picosecond pulse length.

---

* - corresponding author; e-mail : gam110@rsphysse.anu.edu.au

## I. INTRODUCTION

Phase transformations induced by femtosecond laser pulses in metals and semiconductors have attracted significant attention for two main reasons: the quest for improved understanding of the microscopic nature of phase transitions[1-5] on femtosecond time scale and nanometer space scale; and the development of applications such as, for example, an all-optical switch in photonics.[6-9]

The pump-probe technique with subpicosecond laser pulses was broadly used for studying these transformations during the last twenty years. The transient reflectivity of the laser-excited sample was measured with a single optical probe technique with time resolution up to 100 fs while the use of two optical probes allowed one to recover the real and imaginary parts of the full transient dielectric function of a material undergoing phase transition.[10,11] The reflectivities measured by the optical probes directly follow the changes in the electronic properties of the laser-excited matter. In order to observe the structural changes in a lattice on a picosecond time scale the non-linear probe[12] and ultra-short x-ray probes[1,13] were explored together with the optical probe. Several new phenomena were uncovered in these studies. The electron temperature in a solid rises swiftly under the femtosecond laser excitation while the lattice remains cold. The electrons transfer the absorbed laser energy to the lattice via the electron-phonon collisions and heat conduction. The structural changes that occur during the period shorter than the electron-to-lattice energy transfer time are referred to as non-thermal phase transition (or, coherent displacement of atoms[1]). The phase transformation that develops after the electron-to-lattice temperature equilibration is referred to as ultra-fast thermal transition.

The natural question arises: What is the phase state of a solid excited by ultra-short pulse action? The numerous reflectivity measurements with a single probe suggest that several picoseconds after the excitation, a solid is melted because the maximum reflectivity value corresponds to that of the melted stage. We demonstrate in this paper that the phase transition is not completed in as long as 20 ps after the pulse termination even in the case when the deposited energy density exceeds twofold the equilibrium enthalpy of melting. Moreover, there is no direct experimental evidence if the transient phase-state is a crystal-melt mix or a mixture of two different crystalline phases, possibly unknown ones.

We have recently demonstrated with a single probe technique that the reflectivity of a Gallium film excited by femtosecond laser pulses increases from ~55% to ~80% before recovering to its initial value after ~ 0.1-1 μs.[7] This evolution was attributed to a reversible phase transition from the crystalline α-Gallium to the liquid state because the reflectivity maximum corresponds to that of liquid Gallium. To ensure the process was reversible, the total laser fluence was kept well below the ablation threshold *(F_{pump} << F_{abl})* and the average pulse intensity was below the threshold for plasma formation ($I_{laser} \sim 10^{10}$ W/cm$^2$). We have shown in measurements using a single probe beam that the magnitude and rate of the reflectivity changes as well as the time of recovery to the initial state depend on the excitation conditions and background temperature of the Ga film.[7] However, the transient reflectivity obtained with a single probe beam did not allow us an unambiguous evaluation of the material parameters during the phase transition.

In this paper, we used one pump beam and two identical simultaneous 150-femtosecond probe beams (at 775nm wavelength) set at two different angles of incidence: 12° and 32°. The simultaneous determination of the transient reflectivity for the two probes allowed us to recover the time-resolved real and imaginary parts of the dielectric function *ε(ω,t)* of the Gallium film during the phase transition. One can also work out the volume fraction of liquid phase from the measured dielectric function assuming that the transient state consists from the mixture of crystalline and liquid Gallium. It was remarkable to find that the transformation into another phase never reached 100% during the observation time. We have shown that the main reason for incomplete phase transformation is a small depth of laser-excited Gallium layer comprising 10 to 20 atomic distances. Subsequently, the transient dielectric function allowed us to recover the time dependent electron-phonon effective collision frequency (sometimes referred to as optical or transport frequency) and plasma frequency during the phase transition. The transient values of these parameters for the intermediate state created by the laser were found to be drastically different from those in the equilibrium for the crystalline and liquid phase.

The space and time dependent electron and lattice temperatures in the laser-excited skin layer were reconstructed from the experimental reflectivity data without any *ad hoc* assumptions. Then, the phase transformation history in space and in time was reconstructed on the basis of the heating and cooling processes of the skin layer.

The paper is organized as follows. The experimental set-up, the target structure and diagnostics are described in Section II. The experimentally determined transient reflectivity, dielectric function *ε(t)*, electron-phonon collision rate, plasma frequency, and volume fraction of a new phase are presented in Section III. The heating due to the laser absorption, the electron-phonon energy exchange, and the cooling due to electron heat conduction, all are restored using the experimental data only. The threshold energy density for the phase transition and the temperature dependence of the electron-phonon coupling rate are calculated in Section IV. The non-equilibrium and thermal stages of phase transformation are considered in a restricted space of the skin layer and discussed in Section V, and the conclusions are presented in Section VI.

## II. EXPERIMENTAL SET-UP

The scheme of the experimental set-up is shown in Fig.1. A femtosecond Ti:sapphire laser (Clark-MXR CPA 2001, $\lambda$ = 775 nm , pulse duration $t_p$ = 150 fs, pulse energy $E_p \cong$ 1 mJ, repetition rate 1 kHz) provided the s-polarized pump pulse and the two p-polarized probes. The target consisted of a 1–2 μm thick Ga-film deposited on a 1 mm thick SiO$_2$ substrate. The pump was normally incident onto the target from the silica side, as were the probes, that were set at angles of incidence of 12$^°$ and 32$^°$ respectively. The probe beams were tightly focused using $f$ = 100 mm lenses so that the region probed corresponded to the most uniform central 5% of the pump spot. The pump and probe beam dimensions, and the beam overlap were monitored using a CCD camera and microscope imaging system. The pump and probe beam areas were $S_{pump} \approx$ 6.1×10$^{-4}$ cm$^2$ and $S(\phi_1) \approx S(\phi_2) \approx$ 2.6×10$^{-5}$ cm$^2$. Small probe beams were used to minimize degradation of the temporal resolution caused by the different angles with which the three beams irradiated the target surface. The two probe beams had transverse dimensions of $\Delta x(\phi_1)$ = 32 μm, $\Delta x(\phi_2)$ = 34 μm, and this limited the temporal resolution to ~23 fs for probe 1 and ~70 fs for probe 2.

Delay lines placed in the pump and probe optical paths were adjusted to obtain temporal overlap ($t_0$ = 0) between the three beams. This was determined by detecting the onset of the pump induced reflectivity change using the probe beams individually whilst varying the pump-probe delay time. The resolution of the motion controller used to scan the delay of the pump beam relative to the probes was $\Delta x_{min}$ = 2 μm which corresponds to $\Delta t$ = 13 fs, well below the resolution due to the incident angles of the probes. Using 150-fsec laser pulses the real temporal resolution was therefore limited to about 200 fs in these experiments. The pump and probe beams were cross-polarized (s-polarization for the pump and p-polarization for the probes) to minimize the influence of coherent artifacts when the pump and probes overlapped in time, *i.e.* for time delays within a few hundred fs of $t_0$ = 0.

The pump fluence was varied from 0.1 mJ/cm$^2$ to 6 mJ/cm$^2$ using neutral density filters. The probe beam intensities were adjusted to a level at which they induced no detectable changes in the reflectivity. It was also verified that the two probes had no influence on one other. The reflected signals $R_p(\varphi_1 = 12°)$ and $R_p(\varphi_2 = 32°)$ from the Ga-silica interface were collected by lenses and detected by photodiodes. Iris diaphragms were positioned in the focal planes of the lenses to filter the specular reflection from any scattering coming from the target. The reflectivity of Ga was measured as the ratio of the reflected beam intensity relative to the incoming beams and corrected for the reflection losses from the uncoated silica substrate. The signals from the photodiodes were collected by computer providing the evolution of $R_p(\varphi_1)$ and $R_p(\varphi_2)$ as a function of the delay between the pump and the two synchronized probe pulses. Each transmitted value was averaged on a sufficient number of pulses (typically 32 laser pulses per delay point) to minimize the influence of pulse-to-pulse fluctuations of the laser. The resulting accuracy of the reflectivity was ±0.5% at each point. Numerical analysis of the isoreflectance curves for $p$-polarised beams at 12° and 32° angles of incidence plotted in the $\{Re(\varepsilon), Im(\varepsilon)\}$ plane for both crystalline and liquid Gallium suggested that the resulting accuracy to which the real and imaginary parts of the dielectric functions could be determined was ±2%, and ±1% repectively.[14]

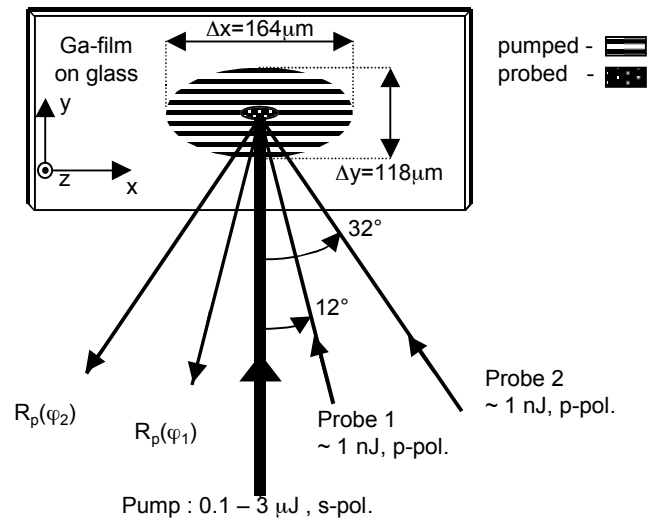

Fig. 1. Arrangement of the pump and two probe beams in the experiments on phase transition induced by femtosecond laser pulses in Gallium films deposited on fused silica substrates.

The Ga-films were deposited using pulsed laser deposition from 6N-purity Ga targets onto Silica substrates at −100°C in a chamber pumped to ~ 2×10⁻⁶ Torr using a Q-switched mode-locked Nd:YAG laser ($\lambda$ = 1.064 μm, $t_{FWHM}$ = 60 ps, intensity on the target $I_0$ ~ 10¹¹ W/cm²) and described in detail elsewhere.[6,7] The crystalline structure of the deposited Ga-film consisted of the α-phase of Gallium that is an elemental metallic molecular crystal containing

Ga$_2$ dimers.[15-20] It has been determined that the covalent bonded Ga$_2$ dimers are oriented almost perpendicular to the film surface.[21,22] The Ga mirrors were mounted on a Peltier cooler to allow the sample temperature to be varied. The temperature at the rear surface of Gallium film was maintained at 16°C during all the experiments, well below the Ga melting point (29.8°C). The target was translated between each run (at varying pump fluence) to provide a fresh Ga-film zone to minimize any effects of cumulative damage, therefore ensuring similar initial operating conditions for all the experiments. Auxiliary experiments were performed to check that the fused silica substrate was insensitive to the level of the pump irradiation (*i.e.* without the Ga-film deposited). No non-reversible changes of the optical properties of the Ga-silica interface were detected for pump fluences up to $F_{pump} \sim 4$ mJ/cm$^2$ and for 10$^6$ laser pulses per spot.

## III. EXPERIMENTAL RESULTS

### A. Transient reflectivity

The reflectivity of the Ga-film, measured at 12° and 32° during the first 25 ps after excitation is plotted in Fig. 2 for three pump fluences. It is clear that the reflectivity grows in two-stages: a rapid rise during the first ~ 2-4 ps was followed by a slower increase over a much longer period. The reflectivity reached its maximum (not shown on Fig. 2) a few hundred picoseconds after the laser excitation.[7] This maximum exceeded the reflectivity data at $t = 25$ ps presented in Fig. 2 by a few percent.

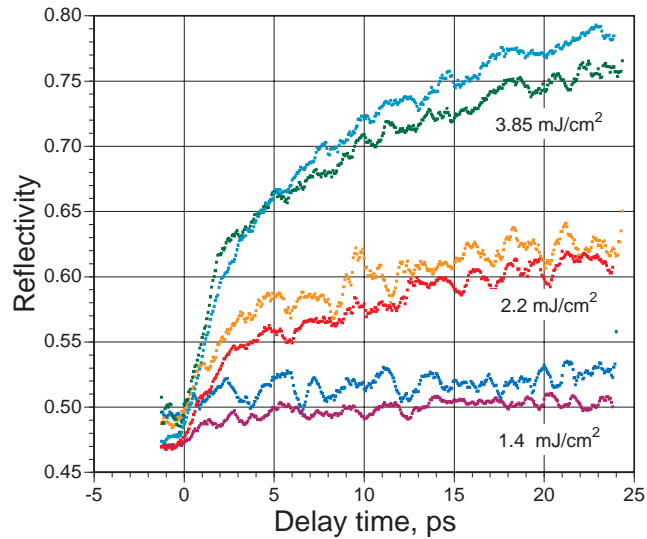

Fig. 2. Probe beams reflectivity change at 12° and 32° in the first 25 ps after the pump pulse at 1.4 mJ/cm$^2$; 2.2 mJ/cm$^2$; and 3.85 mJ/cm$^2$. A two-stage reflectivity rise is clearly seen at all laser fluences.

The maximum reflectivity change at $t = 25$ ps for a pump fluence $F_{pump} \sim 3.85$ mJ/cm$^2$ was in excess of ~50 % for both probes: $R_p(\varphi_1, t = +25$ ps$)/R_p(\varphi_1, t = 0$ ps$) \approx 0.51$ and $R_p(\varphi_2, t = +25$ps$)/R_p(\varphi_2, t = 0$ps$) \approx 0.61$. These values are consistent with the reflectivity changes $R_p(\varphi_1)_{35°C}/R_p(\varphi_1)_{16°C} \approx 0.52$ and $R_p(\varphi_2)_{35°C}/R_p(\varphi_2)_{16°C} \approx 0.57$ measured when the film was heated above its melting point using the Peltier element. The reflectivity changes in such equilibrium conditions represent the transition of Gallium from the α-crystalline phase to a metal-like liquid state. The relative reflectivity change at a fixed delay of 25 ps as a function of pump laser fluence is shown in Fig. 3. The threshold fluence leading to an observable change in reflection was $F_{thrd} \sim 0.5$-0.7 mJcm$^{-2}$, which is in accordance with our previous single-probe experiments.[7]

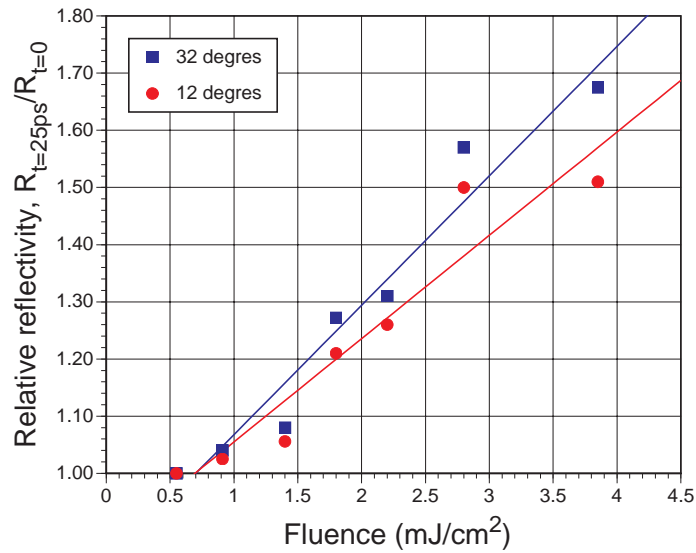

Fig. 3. Relative reflectivity at 25 ps delay time, $R_p(t=+25$ps$)/R_p(t=0$ps$)$, as a function of the pump fluence. Straight lines are linear fits to the data, showing threshold fluence of $0.5 - 0.7$ mJ/cm$^2$.

For $F_{pump} > 4$ mJ/cm$^2$, the sample was clearly damaged by a few minutes of exposure to the pump pulses (~$10^5$ laser shots). This damage threshold was significantly lower than the value of ~40–50 mJ/cm$^2$ obtained using a 30 Hz Ti:sapphire laser.[7] This difference is probably caused by the much higher repetition rate laser used in the present experiments and the use of the low thermal conductivity fused silica substrate that limits the cooling of the Gallium layer between successive laser pulses.

We should note here a regularly observed feature for all the reflectivity measurements with one and with two probes, which is a presence of oscillations with the period of order of picoseconds, and the amplitude well above the experimental error. Such oscillations of reflectivity on the picosecond time scale were observed in a number of experiments.[23,24] These oscillations are related to the generation of (1-0.1) THz optical phonons. The detailed study of this phenomenon is out of scope of the present paper.

### B. Transient dielectric function: Mixture of α-Ga and liquid Ga or intermediate transient phase?

Let us consider as an example, the transient dielectric function determined from the reflectivity measurements (Fig.4a) for intermediate fluence, $F_{pump} \sim 2.2$ mJ/cm$^2$, that is well below the damage threshold.

The reflection coefficient is related to the real, $\varepsilon_1 = \mathrm{Re}(\varepsilon)$, and imaginary, $\varepsilon_2 = \mathrm{Im}(\varepsilon)$, parts of the dielectric function through the Fresnel formulae for $s$- and $p$-polarized light.[25] Thus, two reflectivity values at different angles of incidence form a set of two algebraic equations for two unknowns, $\varepsilon_1$ and $\varepsilon_2$. Numerical solution of these equations as a function of time produces the total complex transient dielectric function of Gallium during the phase transition (see Fig.4a,b). For these calculations, we assumed that polarization of probe beams is unchanged by the Ga-film. The *a posteriori* calculated reflectivity values using the dielectric function obtained in accordance to the above procedure is also shown in Figure 4a. The solutions accurately reproduced the evolution of the reflectivity observed experimentally. Moreover, good agreement between the calculated dielectric function and that known from the literature for Ga in the α-crystalline and liquid states is obtained as shown in Table I.

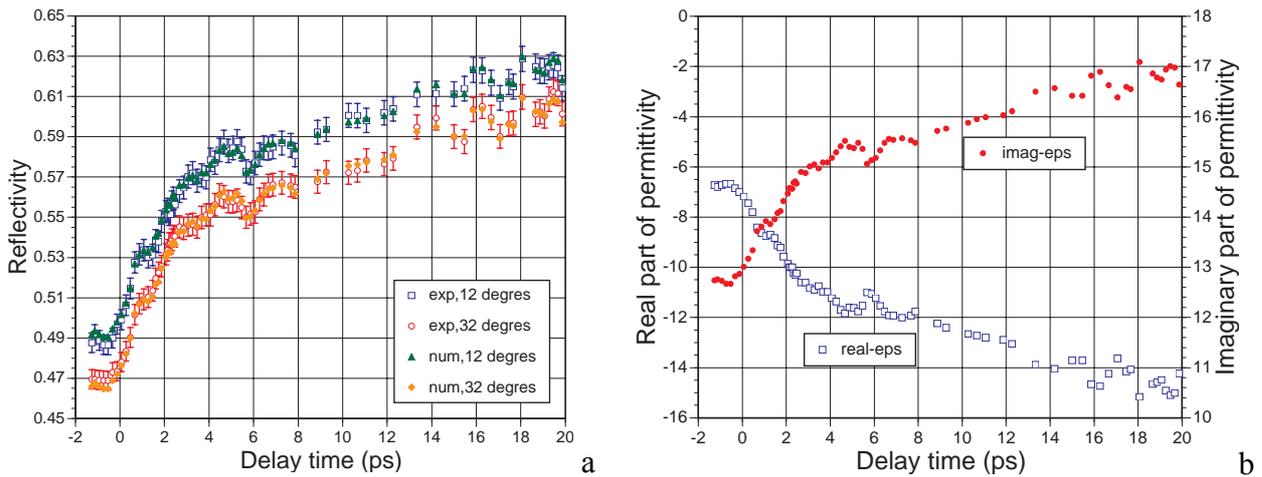

Fig. 4. a) Measured reflectivity (empty symbols) for 12° and 32° at the fluence level 2.2 mJ/cm$^2$, and numerically restored reflectivity (filled symbols) from the recovered dielectric function by inverting the Fresnel formulae, presented as the calculation's accuracy test; b) recovered real (empty squares) and imaginary (dots) parts of the dielectric function.

It is clearly seen from Fig.4b that both real and imaginary parts of the reflectivity lie between the values for crystal and liquid Gallium (see Table I). Thus, the simplest interpretation is that the measured dielectric function, $\varepsilon$, corresponds to that of a mixture of the two equilibrium phases, $\alpha$-Gallium with the dielectric function, $\varepsilon_{\alpha\text{-}Ga}$, and the liquid phase with the dielectric function, $\varepsilon_{liq\text{-}Ga}$.

Table 1. Dielectric function of Ga, tabulated and recovered from the experiments, for $\alpha$-crystalline and liquid Gallium states at $\lambda = 775$ nm (the experimental liquid state values were obtained by melting the film with the Peltier element). The tabulated reference dielectric functions for different Ga states are consistent with the data in Refs. [26-28] and provide independent reflectivity data allowing to check the code accuracy.

| $\varepsilon = \varepsilon_1 + i\varepsilon_2$ | $\alpha$-crystalline state | | liquid state | |
|---|---|---|---|---|
| | $\varepsilon_1$ | $\varepsilon_2$ | $\varepsilon_1$ | $\varepsilon_2$ |
| Tabulated[a] | −7.05 | 12.79 | −66.00 | 34.93 |
| Recovered from experiments | −6.85 | 12.82 | −66.89 | 35.36 |

[a] with courtesy from V. Albanis, University of Southampton, United Kingdom.

We consider the transient state of Gallium as a homogeneous mixture of two phases ignoring the difference of 0.19 g/cm³ in the material densities of the crystalline and liquid phases. We calculate the volume fraction of liquid with the use the Maxwell Garnett formula,[29] that gives correct values for limits of both low and high concentrations, and coincides with an exact Landau formula in the limit of small concentration (up to 40%).[25] Thus, the volume fraction of the liquid phase, $C$, in a bulk material formed a crystalline $\alpha$-Gallium determines the dielectric function of the mixture of phases $\varepsilon_{trans}$ as follows:[29]

$$\varepsilon_{trans} = \varepsilon_{\alpha\text{-}Ga}\left[1 + \frac{3C\left(\varepsilon_{liq\text{-}Ga} - \varepsilon_{\alpha\text{-}Ga}\right)}{\varepsilon_{liq\text{-}Ga} + 2\varepsilon_{\alpha\text{-}Ga} - C\left(\varepsilon_{liq\text{-}Ga} - \varepsilon_{\alpha\text{-}Ga}\right)}\right]. \tag{1}$$

The concentration of liquid Gallium in the transient phase can be calculated by applying this formula to the modulus of the dielectric function $|\varepsilon| = (\varepsilon_1{}^2 + \varepsilon_2{}^2)^{1/2}$, (see Fig.5). It follows from Fig. 5 that the phase transition into a liquid Gallium is not completed during the first 20 ps for the whole range of the pump fluences in these experiments below the damage threshold. At the maximum laser fluence of 2.85 mJ/cm$^2$ the laser energy density deposited into the skin layer is almost two times higher than the equilibrium enthalpy of melting of 479.2 J/cm$^3$ (see Section IV-B below).  We should note, however, that this interpretation of the physical state of the material as a mixture of the two limiting phases is only one of the possibilities.  We cannot exclude the possibility that the material undergoes progressive transformation into an intermediate phase, which is different from liquid or crystalline Gallium as a result of irradiation.

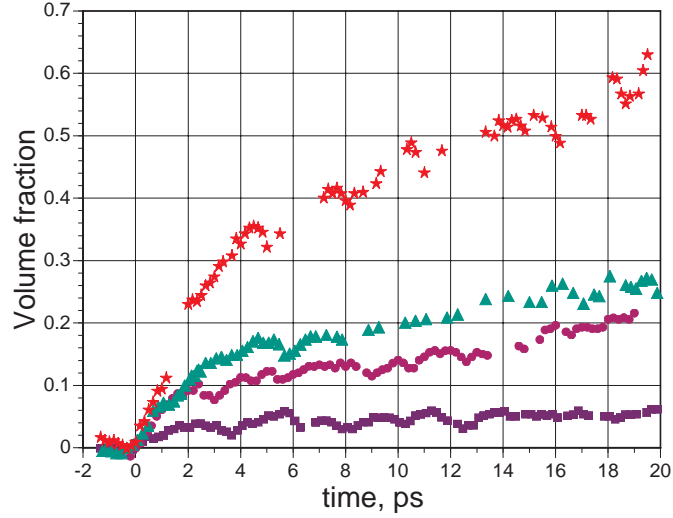

Fig.5.  Volume fraction of the liquid phase in the first 20 ps after the excitation, for the pump fluence 1.4 mJ/cm$^2$ (squares); 1.85 mJ/cm$^2$ (circles); 2.2 mJ/cm$^2$ (triangles); 2.85 mJ/cm$^2$ (stars), calculated using the recovered from experiments transient dielectric function.

C.  Electron-phonon efficient collision frequency and plasma frequency in the mixture of phases

α-Gallium is a molecular crystal with a combination of molecular and metallic characteristics[15-18] while liquid Gallium[19] is a free electron-like metal.  The dielectric function of Gallium in equilibrium for both crystalline and liquid state is well described by the Drude-like form:[26-28]

$$\varepsilon = 1 - \frac{\omega_p^2}{\omega(\omega + i\nu_{opt})} = 1 - \frac{\omega_p^2}{(\omega^2 + \nu_{opt}^2)} + i\frac{\omega_p^2}{(\omega^2 + \nu_{opt}^2)}\frac{\nu_{opt}}{\omega}. \qquad (2a)$$

The plasma frequency, $\omega_p^2/\omega^2$, and the electron-phonon effective collision frequency (optical rate), $\nu_{opt}/\omega$, are both expressed in units of $\omega$, the angular frequency of the probe laser beam ($\omega = 2\pi c/\lambda$) through the real and imaginary parts of the dielectric function:

$$\left(\frac{\omega_p}{\omega}\right)^2 = (1 - \varepsilon_1)\left[1 + \left(\frac{\varepsilon_2}{1 - \varepsilon_1}\right)^2\right]; \qquad \frac{\nu_{opt}}{\omega} = \frac{\varepsilon_2}{1 - \varepsilon_1} \qquad (2b)$$

This two-parametric form describes well the experimentally observed optical properties of both $\alpha$-Gallium,[20,26,27] and liquid Gallium,[26,28] in equilibrium conditions at 775 nm (see Table II). Time-resolved dependences of $\omega^2_p/\omega^2$ and $\nu_{opt}/\omega$ recovered from the reflectivity measurements are presented in Fig. 6.

Table 2.  Plasma frequency $(\omega_p/\omega)^2$ and electron-phonon optical rate $\nu_{opt}/\omega$ for $\alpha$-Ga and liquid Ga for $\lambda = 775$ nm at the temperature below and above the melting point; tabulated[20,26-28] and recovered from the experiments.

| | $(\omega_p/\omega)^2$ | | $\nu_{opt}/\omega$ | |
|---|---|---|---|---|
| | Tabulated | Recovered | Tabulated | Recovered |
| $\alpha$-Gallium | 28.4 | 28.8 | 1.59 | 1.63 |
| Liquid Gallium | 85.2 | 86.3 | 0.521 | 0.521 |

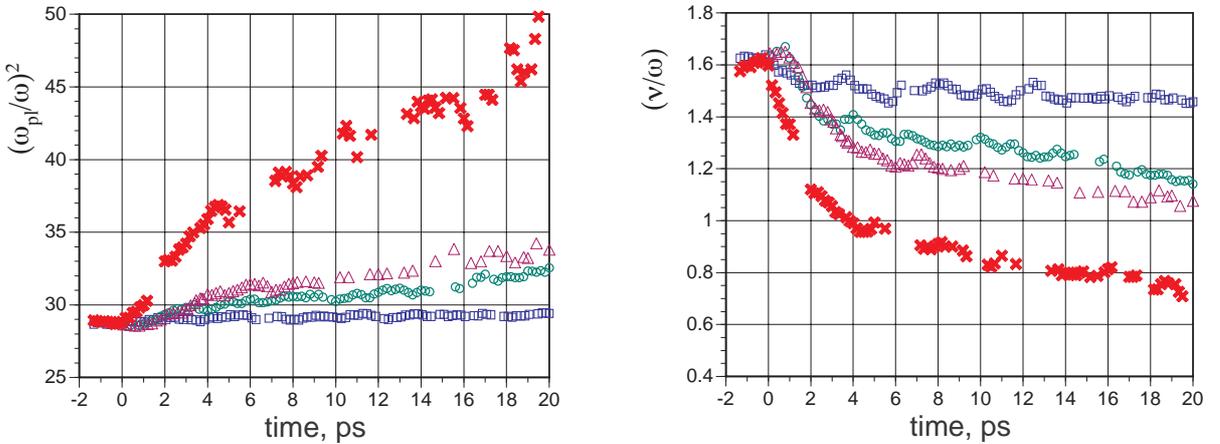

Fig. 6.  Plasma frequency $(\omega_p/\omega)^2$ and electron-phonon collision rate $\nu_{opt}/\omega$ in the first 20 ps after the pump pulse for 1.4 mJ/cm$^2$ (squares); 1.8 mJ/cm$^2$ (circles); 2.2 mJ/cm$^2$ (triangles) and 2.85 mJ/cm$^2$ (crosses) laser fluences.

A close examination of Figure 6 shows that even at the largest deposited energy of 2.85 mJ/cm$^2$ in the graph, being four times larger than the threshold required to observe reflectivity changes, neither the electron-phonon collision rate nor the plasma frequency reach the values for liquid Gallium in equilibrium conditions.  The existence of two time stages with different slopes is also evident in Fig. 6; rapid changes occur during the first 2-4 ps followed by a slower evolution over a period of several tens of picoseconds.

For later analysis, it is convenient to approximate the time dependence of both parameters by power laws:

$$\left(\frac{\omega_p}{\omega}\right)^2 \propto t^{\alpha}; \quad \frac{\nu_{opt}}{\omega} \propto t^{-\beta}; \quad \alpha, \beta > 0. \tag{3}$$

The power dependences for the two stages are distinctly different. The initial fast decrease of the electron-phonon collision rate depends strongly on the amount of the energy deposited. The interpolation shows that power exponent for the time dependence of the electron-phonon collision rate changes within the limits: $\alpha \sim 0.004$–$0.124; \beta \sim 0.03$-$0.23$ for the range of fluences used in the experiments.

### D. Relation between the optical rate and the electron-phonon energy exchange rate

The electron and lattice temperatures both change in all stages of the laser energy absorption, the absorbed energy transfer from the electrons to the lattice, and the lattice cooling by the electron heat conduction. The transformation into a new phase and reverse transition back to the crystalline phase is driven by the electron and lattice temperature distribution, both in time and in space. The measured transient real and imaginary parts of the dielectric function allow one to calculate the electron-phonon effective collision rate, $\nu_{opt}$, (optical, or transport effective frequency) and the plasma frequency.

The electron-to-phonon energy exchange rate $\nu_{en}$ expresses through the experimentally measured optical rate by the following procedure. The conductivity of metals is conventionally calculated for the low values of the adiabaticity parameter $\left(\hbar\omega_D / \varepsilon_F\right)^{1/2} << 1$ (low temperature $T_L, T_e << \varepsilon_F$), by the means of the density functional theory;[31] here $\omega_D$ is the Debye frequency. The optical (transport) relaxation time that defines the conductivity expresses in the theory through the moments of the electron-phonon spectral function as the following:

$$\nu_{opt} \equiv \tau^{-1} \approx 2\lambda\langle\omega\rangle; \quad \omega >> \omega_D; \tag{4}$$

here the bracketed frequency is the first moment of the phonon frequency averaged over the electron-phonon spectral function. The $n$-th moment of the frequency reads:[31,33]

$$\lambda\langle\omega^n\rangle = 2\int_0^\infty d\Omega \left[\frac{\alpha^2 F(\Omega)}{\Omega}\right]\Omega^n. \tag{5}$$

The $\lambda$-function, $\lambda(\omega,T)$, relates to the effective (re-normalized) electron mass, $m^*$. It is expressed through the zero moment of the spectral function. The electron-phonon energy exchange rate (or, the inverse electron-phonon equilibration time $t_{e\text{-}ph}$) was introduced in two-temperature approximation (electrons and phonons are characterized by separate temperatures, $T_e$, $T_L$)[32,33] in the following form:[33]

$$\nu_{en} \equiv t_{e\text{-}ph}^{-1} \approx \frac{3\pi\hbar}{2\varepsilon_F}\lambda\left\langle\omega^2\right\rangle \tag{6}$$

We omitted the multiplier that takes into account the weak dependence of the energy exchange rate on the electron temperature[33] in Eq.(6) because this correction comprises less then 10% at its maximum value ($T_e > T_L$, and $T_e << e_F$). Now, the relation between the optical and energy exchange rate reads:

$$\frac{\nu_{opt}}{\nu_{en}} \approx \frac{4}{3}\frac{\varepsilon_F}{\hbar}\frac{\left\langle\omega\right\rangle}{\left\langle\omega^2\right\rangle} \tag{7}$$

The physical meaning of the bracketed frequency is the average frequency and average of the square of phonon frequency. One can estimate the average phonon frequency in adiabatic approximation near the Debye temperature $T \sim T_D$ as the following:[34]

$$\left\langle\omega\right\rangle \approx \left(\frac{m^*}{M}\right)^{1/2}\frac{\varepsilon_F}{\hbar}; \quad T_L \sim T_D. \tag{8}$$

Thus, the energy exchange rate links to the optical rate through the fundamental characteristics of a material as the following:[34]

$$\nu_{en} \approx \frac{3}{4}\nu_{opt}\left(\frac{m^*}{M}\right)^{1/2}. \tag{9}$$

We adopt this relation for the future calculations, keeping in mind that the time dependence (and therefore, the temperature dependence, as we show below) of the optical rate is obtained from the experiments. The energy exchange rate can be recovered directly from the experimental data. We present the optical data for three different metals: Aluminium,[43] Copper,[43] and Gallium,[20,26-28] in equilibrium conditions at $\lambda = 775$ nm ($\omega = 2.43\times10^{15}$ s$^{-1}$), as examples of these relaxation times in Table 3 below.

Table 3. Optical characteristics of $\alpha$−Ga ($a$-axis), Al, and Cu at 775 nm, and calculated optical and electron-phonon energy rates. The $m*/m_e$ values are from Ref. [35].

| | $R$ | $n$ | $k$ | $\varepsilon'$ | $\varepsilon''$ | $m*/m_e$ | $t_{opt}=\nu_{opt}^{-1}$ | $t_{e-ph}=\nu_{en}^{-1}$ |
|---|---|---|---|---|---|---|---|---|
| $\alpha$−Ga | 0.633 | 2.023 | 3.593 | −8.815 | 14.540 | 1.724 | 0.28 fs | 101 fs |
| Al | 0.88 | 2.6 | 8.42 | −64.136 | 43.784 | 0.676 | 0.61 fs | 221 fs |
| Cu | 0.96 | 0.242 | 4.84 | −23.367 | 2.343 | 0.725 | 4.3 fs | 2.29 ps |

It can be seen from Table 3 that the energy relaxation time depends strongly on the absorption properties of the material. In highly reflective metals like Copper this relaxation time compares to a few picoseconds, while in Gallium it is 100 fs. Note that the electron-phonon energy exchange time in Ga is shorter than the 150 fs laser pulse duration.

Taking the known relaxation rates, $\nu_{opt}$ and $\nu_{en}$, the time history of the electron and lattice temperatures in a solid affected by the short laser pulse can be restored, and then the dependence of the effective electron-phonon collision frequency on temperature can be derived. We follow this procedure below.

Another important parameter, the coefficient of electron heat conduction, $\kappa_e$, is also directly related to the measured electron-phonon collision rate:

$$\kappa_e = C_e\left(T_e\right)n_e\frac{l_e \nu_e}{3} \approx C_e\left(T_e\right)n_e\frac{\nu_F^2}{3\nu_{opt}} \quad ; \qquad (10)$$

here $C_e(T_e)$ is the specific heat for the degenerated electrons, $l_e$ is the electron mean free path, $v_e$ is the electron speed, and $v_F$ is the Fermi velocity ($v_F = 1.92 \times 10^6$ m/s for Ga). Therefore, the transient electron and lattice temperatures in the Gallium layer during the laser-target interaction and after the pulse termination can be calculated without any *ad hoc* assumption and only on the basis of the measured reflectivity data.

## IV. HEATING AND COOLING PROCESSES

### A. Energy absorption and transient temperature in Gallium layer

The laser fluence in all experiments was well below the ablation threshold. Therefore, the density of the target remains essentially unchanged, and the laser-matter interaction occurs through the normal skin effect regime. Since there is no mass or momentum loss, the energy conservation law can be expressed in a two temperature approximation ($T_e$, for electrons, $T_L$, for lattice) by the following set of coupled equations:[32,33]

$$C_e n_e \frac{\partial T_e}{\partial t} = -\kappa_e \nabla T_e - n_e \nu_{en}(T_e - T_L) + \nabla \cdot Q_e$$

$$C_L n_a \frac{\partial T_L}{\partial t} = n_e \nu_{en}(T_e - T_L)$$

(11)

Here $C_e$, $C_L$, $n_e$, $n_a$, respectively, are the electron and lattice specific heat and the density of free electrons and atoms ($n_a = 5.098 \times 10^{22}$ atoms/cm$^3$ for Ga); $Q_e$ is the absorbed laser energy per unit time and per unit volume; $\nu_{en}$ is the electron-phonon energy exchange rate from Eq.(9), and $k_e$ is heat conduction coefficient from Eq. (10). The electron heat conduction dominates in the cooling process, and the phonon heat conduction is neglected. The specific heat of the lattice obeys the conventional Dulong-Petit law $C_L = 3k_B$. The specific heat of the degenerated Fermi electron gas ($T_e << \varepsilon_F$) has a conventional form:[35]

$$C_e = \frac{\pi^2}{2} \frac{k_B^2 T_e}{\varepsilon_F} = \frac{\pi^2 k_B^2 T_e}{m^* v_F^2};$$

(12)

where $m^*$, $\varepsilon_F$ respectively are the electron effective mass and the Fermi energy ($\varepsilon_F = 10.4$ eV).[35] $Q_e$ can be calculated for the normal skin effect interaction mode as:[36]

$$Q_e = \frac{2A}{l_s} I_0(r,t) \exp\left(-\frac{2x}{l_s}\right);$$

(13)

here $A$ is the measured absorption coefficient, $l_s = \frac{c}{\omega k}$ is the skin depth ($k$ is the imaginary part of the refractive index, $c$ is the speed of light in vacuum, $x$ is the distance from the surface into the bulk), and $I_0 = \frac{c}{8\pi}|E_0|^2$ is the incident laser intensity, where $E_0$ is the laser electric field.

The characteristic electron heat conduction time $t_{heat}$, which is defined as a time for the heat wave to propagate a distance equal to the skin depth, could be expressed from Eq.(10-11) as:

$$t_{heat} = \frac{3\nu_{opt} l_s^2}{v_F^2}.$$

(14)

In case of $\alpha$-Gallium, taking $k = 3.64$ for $\lambda = 775$ nm,[20] this time $t_{heat} = 3.75 \times 10^{-12}$ s, which is longer than the pulse duration and the electron-to-lattice energy transfer time estimated above. Therefore, the heating and cooling processes are well separated in time in the experiments.

### B. Electron and lattice temperature during the laser pulse

The transient electron and lattice temperature during the pulse can be calculated in two-temperature approximation (Eq.(11)) using the material parameters recovered from the experiments. The results of the numerical integration of the set of Eq.(11) for a laser pulse with a Gaussian shape in time are presented in Fig. 7 for various laser intensities. The electron temperature rises in the middle of the pulse up to 1000 K ($T_e << \varepsilon_F$) for the highest intensity. The energy transfer from the electrons to the lattice is completed to the end of the laser pulse, $T_e = T_L = T$ at $t = t_{e-ph} \cong t_p$. As the heating and cooling processes are well separated in time, any cooling processes during the laser pulse can be neglected. So, the total energy deposited in the Gallium skin layer just after the end of the pulse ($t = t_p$) $E_{tot}$ can be estimated as the sum of the absorbed laser energy $Q_{abs}$ and the internal energy corresponding to the initial constant temperature of the target $Q_{int}$ ($T_0 = 16°C$ or 289 K):

$$Q_{abs} = \left(C_e n_e + C_L n_a\right)T\left(t_p\right) \approx C_L n_a T\left(t_p\right) = \frac{2AF}{l_s}$$

$$C_L >> C_e \;\; \left(T << \varepsilon_F\right); \qquad F = \int_0^{t_p} I_0(t)dt \qquad\qquad (15)$$

$$E_{tot} = Q_{int} + Q_{abs} = C_L n_a T_0 + \frac{2A}{l_s}F$$

The skin depth in Ga for 775 nm comprises $l_s = 34$ nm. Taking the measured absorption coefficient A $\approx 0.5$, the absorbed energy in the skin layer as a function of the laser fluence can be expressed in the form $E$ [J/cm$^3$] $= 3 \times 10^5 F$ [J/cm$^2$]. Therefore, the observed threshold fluence for the reflectivity change of 0.7 mJ/cm$^2$ corresponds to a deposited energy of 210 J/cm$^3$, which is almost 2.5 times lower than the enthalpy of melting of Gallium in equilibrium conditions, 479.2 J/cm$^3$.[29] The highest fluence in our experiments, 2.85 mJ/cm$^2$, corresponds to an energy density of 855 J/cm$^3$.

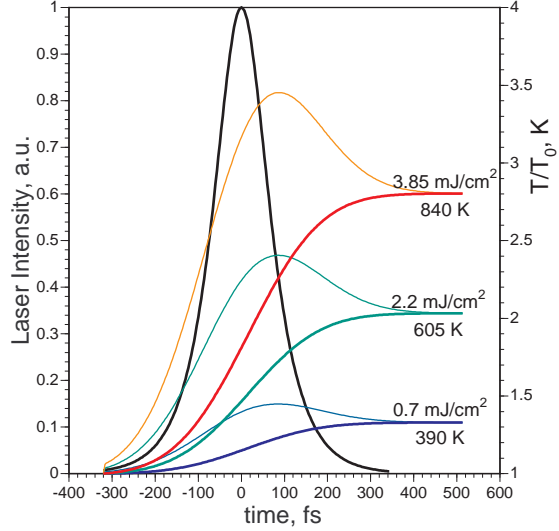

Fig. 7. A Gaussian 150-fs laser pulse, and the electron (thin lines) and lattice (thick lines) temperatures calculated for 0.7 mJ/cm$^2$; 2.2 mJ/cm$^2$; and 3.85 mJ/cm$^2$. The corresponding equilibrated temperatures $T_{max}$ are also shown.

The maximum lattice temperature $T_{max}$ at the silica-Gallium interface after the electron-lattice temperature equilibration can be obtained from Eq.(15):

$$T_{max} = T_0 + \Delta T = T_0 + \frac{2AF}{C_L n_a l_s}. \qquad (16)$$

From this we deduce that the reflectivity starts to change ($F_{thr} \cong 0.7$ mJ/cm$^2$) when the maximum lattice temperature reaches 390 K, well above the melting temperature in equilibrium of 303 K (29.8°C). When the deposited energy density in the Gallium skin layer exceeds the equilibrium enthalpy of melting, which occurs for fluences above 1.65 J/cm$^2$, the lattice temperature rises strongly becoming 605 K at 2.2 J/cm$^2$; 698 K at 2.85 J/cm$^2$; and 840 K at 3.85 mJ/cm$^2$. However, the measured reflectivity values in the experiments (for $t$<20ps) always remained slightly lower than those observed for liquid Gallium and obtained by heating the film up to ~ 309 K (35°C), i.e. above the melting temperature in the equilibrium conditions. As deduced earlier, in the experiments (Figs. 3-6) the phase transition to the liquid state is not complete for at least several tens of picoseconds after the laser pulse in spite of the fact that the electron and lattice temperatures have equilibrated at values far exceeding the melting temperature.

Next consider temperature distribution in the film. In accordance with Eq.(13), the temperature decreases exponentially with the distance from the surface:

$$T = T_0 + \Delta T_{max} \exp\left\{-\frac{2x}{l_s}\right\}. \qquad (17)$$

Therefore, the depth of the layer, $0 < x < x_{melt}$, where the temperature of Gallium exceeds the melting temperature, $T_{melt} < T < T_{max}$, can be expressed as follows:

$$x_{melt} = \frac{l_s}{2} \ln \frac{\Delta T_{max}}{T_{melt} - T_0} \qquad (18)$$

Thus, the thickness of the layer, where the phase transition would be expected to occur, is comparable to the skin depth over the whole range of fluences studied (from 0.7 mJ/cm² to 3.85 mJ/cm²). It comprises $l_s < x_{melt} < 1.8\ l_s$ (34 nm < $x_{melt}$ < 59 nm) and corresponds to a few tens of atomic layers (see Fig.8).

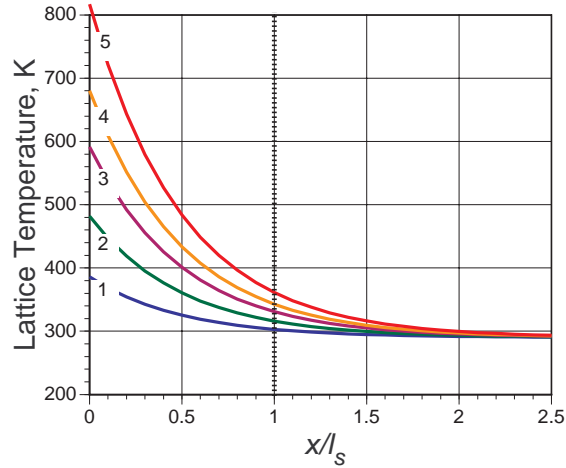

Fig. 8. Maximum lattice temperature distribution with depth in Ga-film at the pump fluence 0.7 mJ/cm² (1); 1.4 mJ/cm² (2); 2.2 mJ/cm² (3); 2.85 mJ/cm² (4); 3.85 mJ/cm² (5).

### C. Cooling by the electron heat conduction after temperature equilibration

At $t > t_{e-ph}$, the energy deposited in the skin-layer is transported away from the layer by means of electron heat conduction in equilibrium conditions, $T_e = T_L$. The lattice specific heat dominates ($C_L \gg C_e$), while the electrons are responsible for the heat transfer. The set of Eq.(11) reduces to the non-linear electron heat conduction equation:

$$\frac{\partial T}{\partial t} = D_0 \frac{\partial}{\partial x} \left( \frac{T}{T_{max}} \right)^{1-n} \frac{\partial T}{\partial x}$$

$$D = \frac{C_e n_e}{C_L n_a} \frac{l_e \nu_e}{3} = D_0 \left( \frac{T}{T_{max}} \right)^{1-n} \qquad (19)$$

$$\nu_{opt} = \nu_0 \left( \frac{T}{T_0} \right)^n$$

$T_{max}$ is the maximum temperature at the moment of electron-phonon equilibration from which the lattice cooling starts, $T_0$ is constant temperature in the bulk of the cold Gallium layer that is controlled by the Peltier element.  We take the diffusion coefficient, $D_0$, as a constant, as the plasma frequency ($\omega^2_p/\omega^2 \propto n_e/m^*$) along with the electron effective mass are slowly varying parameters (see Fig. 6).  The diffusion coefficient expresses as follows:

$$D_0 = \frac{\pi^2}{9} \frac{n_e}{n_a} \frac{k_B T_{max}}{\nu_0 m^*} \left( \frac{T_0}{T_{max}} \right)^n .$$  (20)

The solution for the equation (19) is well known.[37]  One can easily obtain time dependencies for the temperature distribution and the heat wave front propagation using the energy conservation law and Eq.(19) as the following:

$$T(t) = T_{max} \left( \frac{t_0}{t} \right)^{\frac{1}{3-n}}$$  (21)

$$t_0 = \frac{l_s^2}{D_0}; \quad x_{front} = l_s \frac{T_{max}}{T}$$

Thus, a non-linear heat wave with a step-like temperature front propagates inside the bulk of the cold Gallium.  The time dependence of the temperature allows us to recover the time dependence of the electron-phonon collision rate:

$$\nu_{opt} = \nu_0 T^n \propto \left( \frac{t_0}{t} \right)^{\frac{n}{3-n}}$$  (22)

Now, we can relate the experimentally observed time dependence of the electron-phonon collision rate, $\nu_{opt} \sim t^\beta$, to the above formula, to obtain that $n = 3\beta/(1+\beta)$.  The reflectivity, as well as the optical parameters deduced from the reflectivity measurements, changes significantly in the first 2-4 ps after the pulse termination, and is almost constant in next 20 picoseconds.  Thus, $\beta$ changes in the range from 0.03 to 0.23 ($F = 0.7$-$2.85$ mJ/cm$^2$) during the first 2-4 ps, and is almost constant in the remaining part of the observation time 4-20 ps.  That corresponds to changes in $n = 0.087$-$0.56$ in the first 2-4 ps, and $n \approx 0$ after that in the temperature dependence law.  Therefore, the heat conduction law depends on the magnitude of the laser fluence and changes with time during the cooling process in the following way: at $n \approx 0$ it is a non-linear heat conduction with $T \sim t^{-1/3}$, while at $n \to 1$ it tends to the conventional linear process, $T \sim t^{-1/2}$.

The electron-phonon effective collision rate in liquid Gallium under the equilibrium conditions is proportional to temperature, $\nu_{opt} \sim T/T_D$.  The experimental measurements in the temperature range up to 860 K are in close agreement with the linear law.[26,28]  Hence, the temperature dependence of the electron-phonon collision rate tends to that for liquid Ga in equilibrium conditions with the increase of intensity (fluence).  The resulting expression of the recovered non-linear dependence of the electron-phonon collision rate on temperature, $v_{opt} \sim v_0 T^n$ ($n < 1$), is another confirmation of the fact that the crystal-to-liquid phase transition was not completed in the whole range of laser intensities studied here.

The characteristic cooling time, $t_0$, (see Eq.(21)) corresponds to the time for the heat wave to propagate over a distance equal to $l_s$. It is inversely proportional to the deposited energy density:

$$t_0 = \frac{l_s^2}{D_0} \propto T_{max}^{1-n},$$ \hfill (23)

here $T_{max}$ is given by Eq.(16). This time is much longer than the electron-phonon equilibration time, therefore $n \sim 0$. Taking $n_e = 3n_a$, $m^* = m_e$, $v_0 = 2\times10^{15}\,s^{-1}$ for the mixture of phases, and $T_0 = 289K$, one obtains $t_0\,[s] = 4.6\times10^{-8}/T_{max}$ [K]. Thus, the characteristic cooling time ranges from 120 ps (0.7 mJ/cm$^2$) to 55 ps (3.85 mJ/cm$^2$). One can also obtain the time for the temperature in Gallium layer to decrease back to the melting point of $T_{melt} = 303$ K: $t_{melt} = t_0(T_{max}/T_{melt})^2$, ranging from 0.23 ns for 0.7 mJ/cm$^2$ to 0.43 ns (3.85 mJ/cm$^2$). We consider that as the time when the reverse liquid-to-crystal phase transition takes place. This time is in a good agreement with our previous observations with a single probe beam.[7]

## V. DISCUSSION

The change in the material properties induced by the femtosecond laser pulses occurs in a different way during two periods. The first period comprises the time interval shorter than the electron-to-lattice energy transfer time, $t < t_{e\text{-}ph}$. The absorbed energy is deposited to the electrons but the lattice remains cold. The material transition during this time interval is referred to as a "non-thermal – non-equilibrium" phase transition. During the second period, at $t > t_{e\text{-}ph}$, the transition proceeds in conditions for which the electron and lattice temperatures are equilibrated. In this case the phase transition is considered as a thermal one.

The material transformation takes place in a thin layer that comprises only several atomic layers and the thickness of the layer heated above the phase transition temperature varies rapidly in time. We consider, therefore, the implications of the space and time constraints on the kinetics of the phase transformation in such conditions.

### A. The non-equilibrium stage, $T_e >> T_L$

During the non-equilibrium stage that lasts $\sim 100$ fs the electrons gain an average energy in excess of the melting temperature. Any excited electron produces a change in the inter-atomic potential $\Delta U \sim T_e$. This change in turn results in atomic oscillations $\delta R$ around the initial equilibrium position (i.e. the generation of optical phonons):

$$\Delta U \sim \left(\frac{\partial^2 U}{\partial R^2}\right)\delta R^2 \sim \frac{J_i}{d^2}\delta R^2; \qquad (24)$$

Here $d$ and $J_i$ are the inter-atomic distance and the ionisation potential respectively. The average amplitude of atomic oscillation can then be estimated as follows:[34,38]

$$\delta R \approx d\left(\frac{T_e}{J_i}\right)^{1/2}. \qquad (25)$$

According to Ref. [38], the average displacement of an atom for the temperature range of 400-1000 K, typical for melting of the majority of solids, is 0.2-0.25 of the elementary cell radius. The atomic displacement following from Eq.(25) is of the order of a few tenths of Angstrom. The period of these oscillations is of the same order of magnitude as the time necessary to shift an atom into a position corresponding to a new phase. It can be estimated in the framework of the adiabatic approximation.[34] Momentum of an oscillating atom can be estimated from $p\delta R \sim \hbar$. The Virial theorem provides the relation between the oscillation and the kinetic energy $M\omega^2\delta R^2 \sim p^2/2M$. The oscillation period thus could be estimated as follows:

$$t \sim \frac{M\delta R^2}{\hbar}. \qquad (26)$$

This time for Gallium equals to approximately 100 femtoseconds and agrees with similar calculations for Si and GaAs.[4] The process described above can be considered as a "coherent displacement" of lattice layers in the heated skin layer due to electrons excitation. In other words, this is a fast (on the time scale of the laser pulse) loss of long-range (crystalline) order on the space scale of the whole laser-excited layer. There are two other forces that can also lead to a coherent displacement of the lattice during the time before electron-phonon equilibration: the gradient of the electronic pressure, and the ponderomotive force of the laser electric field in the skin layer. The effect of coherent displacement manifests itself in non-linear phenomena such as second-harmonic generation (involving loss of symmetry),[12] or in the change of the x-ray diffraction intensity.[1] In both studies[1,12] the electron-phonon energy exchange time of several ps was significantly longer than the laser pulse duration (~100 fs). Therefore, the effect of non-thermal modification was pronounced.

In the experiment described in this paper, the time for loss-of-long-range-order due to a coherent displacement is comparable to the laser pulse duration and to the electron-to-phonon energy exchange time. Therefore, any non-thermal phase transition could not be observed because of the limited 200 fs resolution in the experiments. Moreover, any change in specular reflection is a manifestation of loss of the short-range order (correlation between close neighbours) and this evolves on the picosecond timescale.

### B. Thermal stage of phase transformation, $T_e = T_{Lattice}$

The crystal lattice heated above the phase transition temperature passes into an unstable state[39] with thermal oscillations of an atom at elevated temperature allowing it to be shifted into a new equilibrium position close to that of a new phase. The transformation into the new phase is seeded, therefore, by this unstable state.

There are two possibilities for developing thermal phase transition in the conditions when the heated skin layer of Gallium is confined between a cold glass substrate and a cold Gallium film that is kept at a constant temperature below the melting point. The first possibility relates to the surface assisted phase transition. The maximum temperature in the heated layer occurs near the Silica-Gallium interface and this suggests that transformation into a new phase would be energetically favored at this interface provided the surface tension between the liquid Gallium and Silica is lower than that for the crystal-liquid Gallium interface.[40] However, the surface tension between the liquid Gallium and Silica is unknown to the best of our knowledge.

On the other side, the formation of liquid seeds in the bulk of a crystal heated from inside while the boundary is kept at the temperature below the melting point is energetically preferred for the bulk melting[39]. Besides, the thickness of the overheated layer increases at high fluences. Consequently, we consider below the formation of seeds of liquid Ga in the bulk of the heated Gallium layer.

The small seeds of the new phase are created in the overheated layer due to lattice fluctuations[39]. These seeds are, however, unstable structures because the formation of an interface between the two phases requires extra energy to overcome the surface tension at that interface. Hence seeds with a size less than a critical value will decay back into the initial phase, whilst seeds with a size exceeding a minimum critical radius will grow rapidly driving transformation of the bulk into the new phase. The critical radius of a seed can be expressed through the temperature of the overheated layer, $T$, as follows:[39]

$$r_{cr} = \frac{2\alpha}{P' - P} \approx \frac{2\alpha}{n_a\left(T - T_{melt}\right)} \tag{27}$$

Here $\alpha$ is the surface tension between the crystal and liquid Gallium ($\alpha$ = 720 dyne/cm),[40] $P'$ is the transient pressure in the skin layer, and $P$ is the pressure corresponding to the melting temperature, $P = n_a T_{melt}$. We neglected a small difference in the densities of liquid and solid Ga ($\rho_{solid}$ = 5.907 g/cm$^3$; $\rho_{liquid}$ = 6.095 g/cm$^3$ near and above the melting point at 29.8°C)[41] in the above calculations. The probability of a seed formation is a strong exponential function of a critical radius[39]:

$$w \propto \exp\left(-\frac{4\pi\alpha r_{cr}^2}{3T}\right). \tag{28}$$

This probability drops sharply with the increase of the critical radius, and the phase transformation process slows down. The lattice temperature grows up during the short period of electron-phonon energy transfer. Afterwards, the temperature slowly decreases ensuring the increase of the critical radius and slowing down the phase transformation process. This is the main reason for the incompleteness of the phase transformation during the femtosecond laser pulse excitation. Therefore the minimum critical radius of a liquid seed corresponds to the maximum lattice temperature just after the pulse termination. The minimum radius decreases from 23.4 nm to 4.4 nm as the fluence increases from 0.7 mJ/cm$^2$ to 2.85 mJ/cm$^2$ (Fig.9). It is clear that the phase transformation at F < 0.5 mJ/cm$^2$ is impossible even if temperature exceeds that for melting because the radius of a critical seed is comparable to the thickness of the layer with $T > T_{melt}$. Similarly, the phase transformation is impossible at the cooling stage when a seed size is comparable to the thickness of the heated layer. As it follows from Eq.(27), the critical radius coincides with the skin depth when the lattice temperature is still 60 K above the melting temperature.

Let us now estimate the seed formation time. We consider the seed formation process as an isotropic atom-to-atom attachment to a seed nucleus with a characteristic thermal velocity $v_{th} = (2T/M)^{1/2} \sim 10^5$ cm/s. Then, the time required for a critical seed to be formed can be estimated as follows:

$$t_{form} \approx \frac{r_{cr}}{3v_{th}}. \tag{29}$$

The time for the formation of a seed with the critical radius of 4 nm ( F = 2.85 mJ/cm$^2$ ) is ~ 4 ps. The optical response builds up at the depth of $c/\omega_{pe}$ that comprises ~ 10$^{-6}$ cm for Gallium. Therefore formation of liquid seeds of several nanometers in size can be qualitatively associated with the abrupt reflectivity changes observed at this fluence in the first 2-4 ps after the laser pulse.

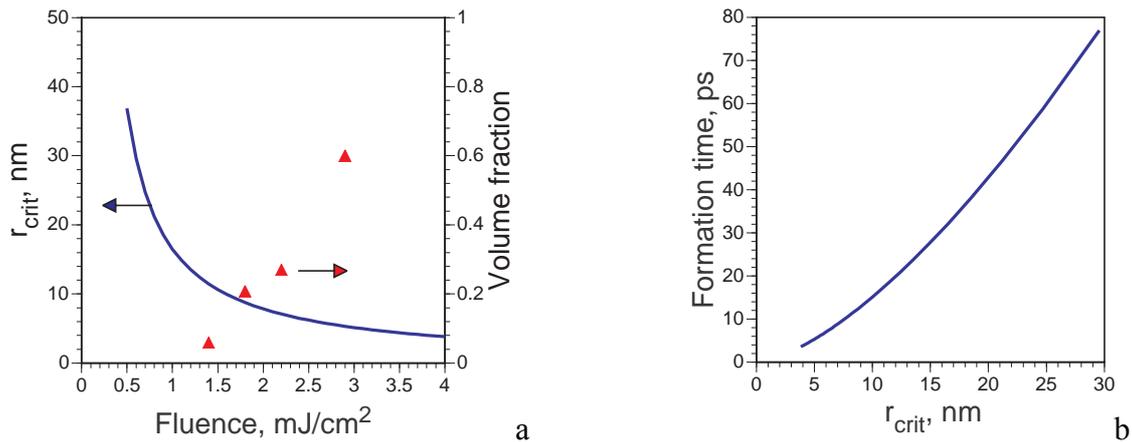

Fig. 9. a) – The calculated critical radius of seeds and volume fraction of a new phase recovered from the experiments *vs* pump fluence; b) – time for seed formation with the critical radius. Note that the critical radius is larger than the skin layer at $F < 0.5$ mJ/cm$^2$, so the phase transition does not occur.

It is also worth noting that during femtosecond excitation of any material, the pressure inside the skin layer increases along with the temperature. In the Gallium skin layer, the pressure varies from 3 to 5 kbar for fluences in the range 0.7 mJ/cm$^2 < F < 2.85$ mJ/cm$^2$. The melting temperature of Ga(I) decreases with increasing pressure: at 1.2 GPa (12 kbar) the melting temperature has dropped to 293 K (room temperature).[42] These changes could affect the dynamics of the phase transition although the change in melting temperature is small compared to the overheating induced by the laser in the interaction region.

## VI. CONCLUSION

In this paper, the reversible phase transition in a Gallium film irradiated by femtosecond laser pulses has been studied by measuring its transient reflectivity with one pump and two identical simultaneous femtosecond probes set at two different angles. These measurements allowed complete determination of the real and imaginary parts of the transient dielectric function with ~ 200 fs time resolution and from these the electron-phonon effective collision rate (optical or transport frequency) and heat conduction coefficient for the first 25ps after irradiation. The time-dependent electron and lattice temperatures in the layer undergoing the phase transition were then determined. The time history of the phase transition in Gallium induced by femtosecond laser pulses was reconstructed on the basis of the experimentally measured transient effective electron-phonon collision frequency in the first 25 ps after the excitation with ~200 fs time resolution.

The main results of this study are the following:

- femtosecond laser excited $\alpha$-Ga transforms into a phase state whose optical properties are intermediate between those for the crystal and for the liquid Gallium. This state is a most likely a coexistence of different phases of Gallium, possibly $\alpha$-Gallium and liquid Gallium.

- The main reason for incomplete phase transformation after the femtosecond laser pulse excitation is the limited thickness of the laser-excited layer. At the energy threshold for the reflectivity change the size of a seed of a new phase is comparable to the skin depth. Correspondingly, the phase transformation terminates during the cooling stage after the laser pulse when the seed size is approaches to the thickness of the excited zone. This makes the laser-excited phase transition drastically different from the phase transformation in a bulk solid in the conditions of thermodynamic equilibrium.

- The experimentally determined threshold for the reflectivity changes, 0.5-0.7 mJ/cm$^2$, corresponds to the absorbed energy density two times lower than the equilibrium enthalpy of melting for Gallium.

- The electron-phonon collision rate and the heat conduction coefficients recovered from the experimental data are transient non-linear functions of temperature, and are drastically different from those in equilibrium.

- The first sharp rise in reflectivity in a few ps after the excitation corresponds to the fast growth of seeds of new phase whilst the electron heat conduction is negligible. The following slow increase in reflectivity corresponds to the conditions when the heat transfer dominates. The lattice cooling through the electron heat conduction slows down, and than terminates the solid-to-liquid phase transformation on 100-200 ps timescale.[7]

- The measured changes in the linear (specular) reflection occurs on the timescale of several picoseconds or greater. The phase transition therefore proceeds in the thermal mode.

The non-thermal stage of the phase transition or, the coherent displacement, takes place during the time comparable to the 150-fs pump duration and could not be resolved in these experiments with 200-fs resolution presented here. The coherent displacement of the lattice excited by the laser pulse on a time-scale shorter than the electron-phonon energy exchange time (< 100 fs), can be studied in future experiments combining ultra-short optical and X-ray probing[1] with time resolution better than 100 fs. These experiments would lead to a precise characterization of the non-equilibrium stage of the laser-excited phase transition.

## ACKNOWLEDGMENTS

The authors are grateful to N. I. Zheludev for a number of useful discussions and suggestions at the early stage of the experiments, and to V. Albanis for the set of data on optical properties of Gallium. O.U. gratefully acknowledges Australian Research Council's financial support through the ARC-IREX scheme (Grant n° X00106527 "Ultrafast laser ablation and deposition of thin films").